\documentclass[a4paper,11pt]{article}
\usepackage{pos}
\usepackage{hyperref} 

\title{The Galactic TeV sky: sources or diffuse emission?}
\author*[a]{Kathrin Egberts}
\author[a]{Constantin Steppa}
\author[a]{Karol Pawel Peters}

\affiliation[a]{
Universit\"at Potsdam, Institut f\"ur Physik und Astronomie, Campus
Golm, Haus 28,\\ Karl-Liebknecht-Str. 24/25, 14476 Potsdam-Golm, Germany}

\emailAdd{kathrin.egberts@uni-potsdam.de}
\abstract{Gamma-ray observations have recently shifted the focus to higher and higher energies, with capable ground-based instruments enabling measurements in the TeV to PeV domain. While a clear prevalence of diffuse emission is observed in the GeV sky, energy-dependent cosmic-ray transport suggests a reversal of this hierarchy at higher energies. Measurements, however, are at strife regarding this question. While imaging atmospheric Cherenkov telescopes (IACTs) see a source-dominated Galactic plane, air-shower particle detectors (ASPDs) report a dominance of diffuse emission. Reconciling these claims might require a closer look at the involved instrument limitations: IACTs have a small field of view, resulting in poorer performance for large-scale emission due to the applied background subtraction technique. ASPDs have reduced resolution capabilities, resulting in unresolved sources contributing to the measurable diffuse emission signal. 
Here we contribute to this controversy by investigating the amount of unresolved sources in current TeV measurements in a population synthesis approach and discuss the unique capabilities for high-resolution diffuse-emission measurements with IACTs and their possibilities for overcoming their background limitations.}

\FullConference{%
  7th Heidelberg International Symposium on High-Energy Gamma-Ray Astronomy (Gamma2022)\\
  4-8 July 2022\\
  Barcelona, Spain\\}

\begin{document}
\maketitle

\section{Introduction}
\begin{figure}
\centering
  \includegraphics[width=.9\textwidth]{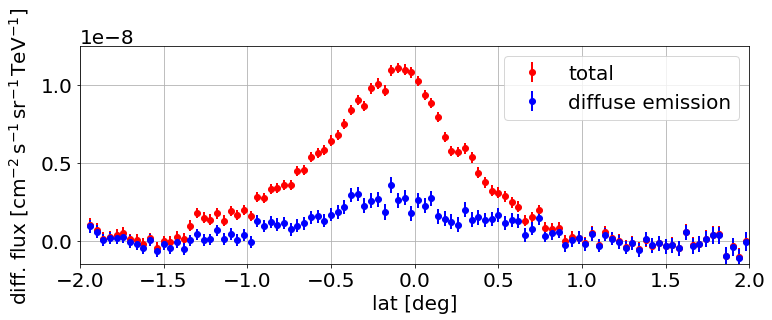}
  \caption{The latitude profile of the total and the diffuse emission measured by the H.E.S.S. telesopes \cite{HESS_diffuse}. The emission is accumulated over a longitude range of $-75^\circ<l<60^\circ$. Given is the differential flux at 1~TeV for the total emission in red and the emission from regions without significantly detected $\gamma$-ray sources in blue. }
\label{FIG:HESSDiffuse}
\end{figure}

\begin{figure}
\centering
  \includegraphics[width=.7\textwidth]{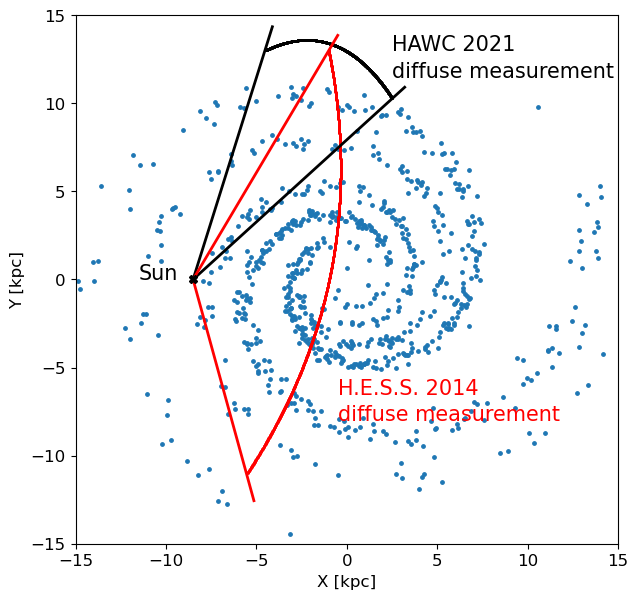}
  \caption{On top of a simulation of the source population based on the mSp4 model \cite{Steppa}, the ranges of the HAWC and H.E.S.S. measurements are shown. HAWC data cover a longitude range of $43^\circ<l<73^\circ$, while H.E.S.S. measurements extend in a longitude range of $-75^\circ<l<60^\circ$.}
\label{FIG:MilkyWay}
\end{figure}

Galactic $\gamma$-ray emission at TeV energies sheds light on the most energetic cosmic-ray processes in the Milky Way. While at GeV energies the Galactic sky is dominated by the large-scale diffuse emission, it is expected that this situation is inverted at some point due to the steeper spectrum of the diffuse emission compared to average source spectra. At TeV energies, $\gamma$-ray measurements are bound to ground-based observations due to the low fluxes and therefore large required detector areas. Two complementary techniques have been established for these measurements: air-shower particle detectors (ASPDs) measure directly the particles produced in the $\gamma$-ray initiated air shower (HAWC \cite{HAWC_catalog}, ARGO-YBJ \cite{Argo-YBJ}, LHAASO \cite{LHAASO}), while imaging atmospheric Cherenkov telescopes (IACTs) detect the Cherenkov light emitted by air-shower particles (e.g. H.E.S.S. \cite{HESS_catalog}). The different techniques grant for different instrument characteristics, namely a large field of view and duty cycle at the expense of a reduced angular resolution for ASPDs on the one hand and a small field of view of few degrees and very limited duty cycle endowed with an angular resolution of smaller than $0.1^\circ$ for IACTs on the other hand. The Galactic sky has been observed with both techniques, reporting the diffuse emission to take up a fraction of $72.7\%$ for $|b| < 2^\circ$ ($76.1\%$ for $|b|<4^\circ$) in the HAWC measurement \cite{HAWC_diffuse} and $25\%$ in the H.E.S.S. measurement \cite{HESS_diffuse}, shown in Fig.~\ref{FIG:HESSDiffuse}. This obvious contradiction is somewhat mitigated by the fact that these measurements are restricted to their accessible sky, depending on the location of the respective instruments. In the Northern hemisphere, observations are limited to the (outer) Eastern parts of the Galaxy (e.g. HAWC, ARGO-YBJ, LHAASO), while only the Southern hemisphere grants the view on the entire inner Galaxy (H.E.S.S.). This situation is visualised in Fig.~\ref{FIG:MilkyWay}. A proper comparison should use the region of overlap rather than the entire measurement. However, the striking disagreement does not necessarily need to be a consequence of the different skies probed but can also be attributed to the instrument characteristics and analysis techniques, as has been demonstrated in \cite{Armelle} in the comparison of source analyses in the region of overlap of the two experiments. 
More precisely, there are two major effects that influence the measured level of diffuse emission: the fraction of unresolved $\gamma$-ray sources in the measured diffuse signal and the effect of the charged-cosmic-ray background subtraction, each of which biases the diffuse measurements of the instruments depending on their characteristic weaknesses.
\section{Unresolved Sources in Diffuse-Emission Measurements}
\begin{figure}
\centering
  \includegraphics[width=.7\textwidth]{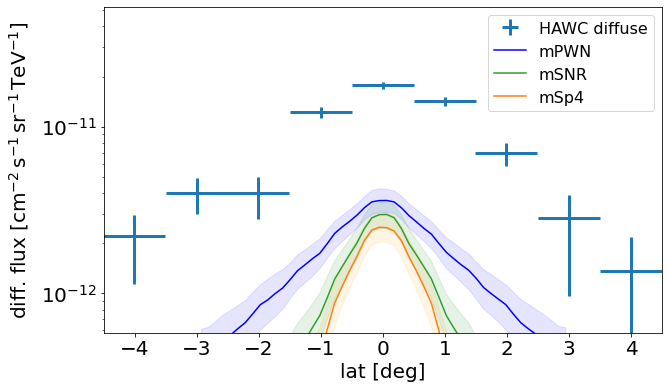}
  \caption{The preliminary latitude profiles of the HAWC diffuse measurement at 7~TeV in the longitude range $43^\circ<l<73^\circ$ \cite{HAWC_diffuse} together with the estimated contribution of unresolved sources for the mSNR, mPWN, and mSp4 models. Models are shown as median derived from 3000 simulated populations, error bands depict the interquartile ranges.}

\label{FIG:LatProfiles}
\end{figure}

  In view of finite instrument sensitivities a measurement of diffuse emission is bound to contain a contribution of $\gamma$-ray sources that are individually too weak to be detected significantly but cumulatively still contribute to the measurable diffuse signal. This contribution of unresolved sources can be expected to play a larger role at higher energies. It depends crucially on the characteristics of the instrument, in particular its overall sensitivity and angular resolution, but also on the details of the analysis like observation time and in particular for pointed instruments on the pointing strategy, i.e. the spatially dependent exposure. To estimate the amount of unresolved sources in the diffuse measurement reported in \cite{HAWC_diffuse}, we follow a population synthesis approach as described in \cite{Steppa}: Based on four of the models for the population of Galactic sources described there, namely two azimuthally-symmetric distributions following the SNR and PWN distribution, mSNR and mPWN, and two models exhibiting spiral arms, one following the matter density, mSp4, and one the free electron density, mFE, (for properties and references of the models see \cite{Steppa}) we use the HAWC exposure to derive longitude and latitude profiles of the sources that are detectable und the ones that are unresolved in the HAWC diffuse emission measurement \cite{HAWC_diffuse}. For this purpose we use the HAWC sensitivity as presented in the 3HWC \cite{HAWC_catalog} as a function of declination and translate it using the quoted observation time of 1347 days in \cite{HAWC_diffuse} to a two-dimensional exposure in longitude and latitude. As the models of \cite{Steppa} only quote fluxes at 1~TeV, we convert this to a 7~TeV flux assuming the average spectral index of the HGPS, 2.3. 3000 source populations are simulated for every model (mSNR, mPWN, mSp4, mFE). Folding in the HAWC point spread function taken from \cite{HAWCwebpage}, we derive the fraction of resolved and unresolved sources by applying the derived detection criterion to the simulations and derive their longitude and latitude profiles as median with interquartile ranges. For verification purposes, the profiles of the detectable sources are compared with the data of the HAWC catalog. Both longitude and latitude profiles give consistent results for the models mSNR and mPWN within the relatively large statistical scatter of the simulations defined by their interquartile ranges. The model mFE exhibits a latitude profile that is too broad to be compatible with HAWC data, for mSp4 the longitude profile exhibits too small a contribution towards the outer Galaxy. The contribution of unresolved sources is presented in Fig. \ref{FIG:LatProfiles} for mSNR, mPWN, and mSp4, together with the preliminary HAWC diffuse measurement. The fraction of unresolved sources for these models lies at around 10\% of the total diffuse emission, rendering its contribution sub-dominant for the investigated models.

\section{Background Estimation for Diffuse-Emission Measurements}
Another effect that influences large-scale emission in particular for small field-of-view instruments is the treatment of the charged cosmic-ray background. With large-scale diffuse emission being a small signal on top of a vastly dominating charged cosmic-ray background, the in-depth understanding of this background is crucial for a reliable measurement. Customarily, this background has been estimated from ``\texttt{OFF} regions'' without $\gamma$-ray emission from within the field of view, which reduces the systematics related to atmospheric conditions and the state of the instrument. This method, which was also used in an adapted form (using generous exclusion of \texttt{OFF} regions within the latitude range of $\pm 1.2^\circ$) for the H.E.S.S. diffuse measurement, can lead to a reduction and distortion of the measurable signal because residual emission in the \texttt{OFF} regions results in an overestimation of the actual background. In \cite{HESS_diffuse} this effect was estimated to be of a level of 30\% for a Gaussian of width of $2^\circ$ (95\% for a Gaussian of width of $20^\circ$). Recent developments have brought forward the use of likelihood fitting methods \cite{Lars} that use background templates derived from extragalactic observations. While these methods reduce the level of distortion, they come with systematics related to varying atmospheric and instrument conditions as well as related to properties intrinsic to the observed sky like the night sky background. Also these methods still rely on a fitting procedure %to determine the accurate level of background
in a region of no $\gamma$-ray emission within the field of view (rehashing the aforementioned problem of \texttt{OFF} regions). Nevertheless, the optimisation of such templates is the only way to overcome the challenges of the charged cosmic-ray background for diffuse emission measurements with small field-of-view instruments. Currently under development is an approach of deriving the background templates from simulations. Based on simulations custom-tailored to each individual observation run \cite{Holler} this method has the potential to provide an absolute measurement of diffuse emssion significantly extending the instrument field of view.

\section{Discussion and Conclusion}
The discrepancy between ASPD and IACT measurements in the fraction of diffuse emission can, independently of the different regions probed by the measurements, naturally be attributed to the systematics caused by the shortcomings of the respective instruments, namely limitations in the resolution of $\gamma$-ray sources for ASPDs and artificial reduction and distortion of the signal due to background subtraction for IACTs. Here we show that the amount of unresolved $\gamma$-ray sources in the diffuse emission measurement performed by HAWC is, for the $\gamma$-ray source population models investigated, with around $10\%$ negligible. However, it has to be noted that out of the four models investigated (all of them have shown to be compatible with the data of the H.E.S.S. Galactic source catalog), only the two azimuth-symmetric ones are not in tension with the observed HAWC source distribution, while both models featuring more realistic spiral-arm structures are excluded. In particular the exclusion of model mSp4, which shows a very good description of H.E.S.S. data and has been treated as a reference model in \cite{Steppa}, due to lacking sources at large longitude values suggests that the description of the outer part of the Galaxy, which the HAWC data probes, is incomplete. A natural extension of the mSp4 model could be the inclusion of the local arm (see e.g. \cite{Reid}), which is not considered in the current model but passes through the HAWC-observed sky and could make an articulated difference. The final assessment of the contribution of unresolved sources in the HAWC data would therefore require the inclusion of this feature.\\
While it is not excluded that the true amount of diffuse emission in ASPD measurements is reduced due to a fraction of unresolved $\gamma$-ray sources, the inverse scenario of an increase in diffuse emission in IACT measurements is also likely from the perspective of background-subtraction issues. Here, more advanced techniques currently under development bear the chance of a better handle of the charged cosmic-ray background and thus a more realistic approach to the true level of diffuse emission. Given the limited field of view of H.E.S.S., such measurements should ideally not require a normalisation from within the field of view to avoid any artificial reduction of the signal. The possibilities implied with such high-resolution IACT measurements are exemplified in the level of detail visible in the H.E.S.S. latitude profile of Fig.~\ref{FIG:HESSDiffuse}. If measurements by Fermi-LAT, HAWC, and ARGO-YBJ, or simply the gas distribution in the Galaxy are indicative for the width of the latitude distribution of diffuse emission, the increased field of view of the forthcoming Cherenkov Telescope Array Observatory could already make a crucial difference and improve IACT diffuse-emission measurements irrespective of the expected boost in sensitivity.

\end{document}